\documentclass[12pt]{iopart}
\usepackage{graphicx}
\usepackage{color}

\usepackage{color}
\definecolor{dblue}{rgb}{0,0,0.75}
\definecolor{dred}{rgb}{0.6,0,0}
\definecolor{dgreen}{rgb}{0,0.5,0}

\begin{document}

\title{Rewiring the network. What helps an innovation to diffuse?}

\author{Katarzyna Sznajd-Weron$^1$, Janusz Szwabi\'nski$^{2,3}$, Rafa{\l} Weron$^4$, Tomasz Weron$^{3,5}$}

\address{$^1$ Institute of Physics, Wroc{\l}aw University of Technology, Poland\\ 
$^2$ Institute of Theoretical Physics, University of Wroc{\l}aw, Poland\\ 
$^3$ UNESCO Chair of Interdisciplinary Studies, University of Wroc{\l}aw, Poland\\
$^4$ Institute of Organization and Management, Wroc{\l}aw University of Technology, Poland\\
$^5$ High School LO14 in Wroc{\l}aw, Poland}

\begin{abstract}
A fundamental question related to innovation diffusion is how the social network structure influences the process. Empirical evidence regarding real-world influence networks is very limited. On the other hand, agent-based modeling literature reports different and at times seemingly contradictory results. In this paper we study innovation diffusion processes for a range of Watts-Strogatz networks in an attempt to shed more light on this problem. Using the so-called Sznajd model as the backbone of opinion dynamics, we find that the published results are in fact consistent and allow to predict the role of network topology in various situations. In particular, the diffusion of innovation is easier on more regular graphs, i.e. with a higher clustering coefficient. Moreover, in the case of uncertainty -- which is particularly high for innovations connected to public health programs or ecological campaigns -- a more clustered network will help the diffusion. On the other hand, when social influence is less important (i.e. in the case of perfect information), a shorter path will help the innovation to spread in the society and -- as a result -- the diffusion will be easiest on a random graph.
\end{abstract}
%\pacs{89.20.-a, 	%Interdisciplinary applications of physics
%89.65.-s, %Social systems
%89.65.Gh,	%Economics; econophysics, financial markets, business and management
%89.75.-k 	%Complex systems
%89.75.Fb, %Structures and organization in complex systems
%89.75.Hc 	%Networks and genealogical trees 
\vspace{2pc}

\maketitle

\section{Introduction}

Diffusion of innovation, defined as a process in which an innovation is communicated through certain channels over time among the members of a social system  \cite{rog:03}, is not only a fascinating but also an extremely important topic of research in many disciplines, including economy, sociology, anthropology and medicine. It is important to realize that an innovation is not necessarily a new technology or product. It can be also any new, or at least perceived as new,  idea or practice like, for instance, proecological or healthy behavior.
What has been observed already in the 1960s by Rogers is that although the diffusion of innovation may concern a very broad spectrum of phenomena there are some universal features \cite{rog:03}. Most notably they include the S-shaped curve of adopted, the existence of a critical mass of adopters and the multi-stage character of the innovation-diffusion process:
\begin{enumerate}
\item
the knowledge stage (gain knowledge of an innovation), 
\item
the persuasion stage (form an attitude/opinion towards it), 
\item
the decision stage (decide to adopt or reject it),
\item
the implementation stage (implement it) and
\item
the confirmation stage (confirm the decision).
\end{enumerate}

Quite likely it is because of this multi-stage nature of the process and the usually high level of uncertainty associated with the introduction of many innovations, social norms and social influence play a fundamental role in the diffusion of innovation. It has been shown empirically that in some cases, for example, in case of environmental behaviors, social norms and influence have a greater impact than other non-normative motivations like protection of the environment, benefiting the society or even saving money \cite{nol:sch:cia:gol:gri:08,alc:11,ayr:ras:shi:13}, see also the discussion and references in \cite{kow:mac:s-w:wer:13}. Interestingly, at the same time consumers (agents) are not aware of the power of social influence -- \emph{because others are doing it} was judged to be the least important reason at the self-reported motivation stage \cite{nol:sch:cia:gol:gri:08}. 

The critical relevance of social influence in the diffusion of innovations has been recognized for some time now. However, traditional aggregate (analytical) models -- like the Bass model -- are not behaviorally based \cite{gol:lib:sol:sta:00} and therefore cannot reproduce the complexity of real-world diffusion patterns. No wonder that \emph{agent-based modeling}, which -- from the statistical physics point of view -- is nothing else than a microscopic approach, was employed by several researchers already in the 1980s; for a historical overview see \cite{kie:gun:stu:wak:12}. With the development of numerical techniques and computational power this trend has has picked up momentum over the years and is increasingly being used in this field \cite{gar:05,gar:jag:11,prz:s-w:wer:13}.

A fundamental question regarding diffusion of innovation is how the social network structure influences the process. However, as noted in \cite{Per:Mul:Mah:10}, research on the role of network structure is still in its infancy, mainly because of the lack of empirical data.
First of all, although considerable amount of work has been dedicated to measure the properties of large-scale networks, only a few researchers have attempted to study word-of-mouth influence empirically. As a result, evidence regarding real-world influence networks is limited \cite{wat:dod:07}. Secondly, it is highly unclear if and how the structure of the social network influences the diffusion of innovation. In the published studies one can find different and at times seemingly contradictory results. We should stress here that these results relate to the behavior of agent-based models of the diffusion of innovation on different networks, not to empirical studies. As noted by Delre et al \cite{Del:Jag:Bij:Jan:10}, it is very difficult to conduct controlled experiments on processes of innovation diffusion but fortunately, simulation models provide a tool to systematically conduct experiments. Although it is hard to disagree with the first part of the statement, one should be very careful in treating computer simulations as a substitute for real experiments. Long-term experience from statistical physics shows that results may strongly depend on model details. Therefore, no wonder that the problem of the role of topology for the diffusion of innovation is still unclear. 

For instance, Kocsis and Kun \cite{koc:kun:08} have shown that the degree polydispersity and long range connections of agents can facilitate, but can also hinder the spreading of new innovations, depending on the amount of advantages provided by the innovation. Lin and Li \cite{lin:li:10} have shown within an agent-based model that the diffusion of knowledge is easier on a small-world network than on a regular network but harder than in the case of a random graph. On the other hand Watts and Dodds \cite{wat:dod:07} and more recently Bohlmann et al. \cite{boh:cal:zha:10} have shown that network effects depend on how we model the social impact. In turn, Pegoretti et al. \cite{peg:ren:mar:12} have shown that with perfect information, the innovation diffuses faster when the network is completely random but with imperfect information, the fastest pace is reached in small-world networks.

In this paper we address not only the question posed by Bohlmann et al. \cite{boh:cal:zha:10}:
\emph{Do the effects of adoption threshold on diffusion cascades differ by network structure?} We try to answer also a broader question: \emph{Which of the properties of the network and how affects the diffusion of innovation?} Of course, we realize that we can address this question only within a particular agent-based model. However, taking into account the results obtained by other researchers for other models, we can speculate on the universality of our results and predict under what circumstances we can expect a different scenario.

The remainder of the paper is structured as follows. In Section \ref{sec:model} we describe and motivate the agent-based model we use in our simulation study. In Section \ref{sec:results} we discuss the results we obtain for different model structures. Finally, in Section \ref{sec:conclusions} we wrap up the results and conclude.

\section{Model}
\label{sec:model}

We consider a set of $N$ agents, each of which may be in one of two states: $S_i=-1$ (not adopted) or $S_i=1$ (adopted), for $i=1,\dots,N$. Following \cite{prz:s-w:wer:13} and \cite{nyc:s-w:13}, we  call these agents \textit{spinsons} (spin+person) to reflect their dichotomous nature originating in spin models of statistical physics and humanly features and interpretation. We should mention that the idea of using binary variables in modeling innovation diffusion is not new \cite{gol:bar:eit:01,gal:vig:05,kin:sol:sta:13,alt:bra:dal:zec:13,lac:rov:11}. In fact Watts and Dodds \cite{wat:dod:07} argue that binary decisions can be applied to a surprisingly wide range of real world situations and are particularly useful for modeling processes in which `adopted' and `not adopted' are natural states. On the other hand, binary variables are certainly a simplification that does not take into account the complexity of human decision making. Therefore, also non-binary characteristics have been considered in the diffusion of innovation literature, for instance, continuous variables to describe social values \cite{def:hue:amb:05} and discrete awareness states \cite{kra:red:vol:11}.

There are different types of social influence a spinson may be exposed to, see \cite{prz:s-w:wer:13} for an overview. However, three of them (and their interplay) are of particular interest for studying the diffusion of innovation:
\begin{itemize}
\item 
\textit{Independence}, understood as unwillingness to yield to group pressure, is a particular type of non-conformity \cite{s-w:tab:tim:11}. Independence introduces indetermination in the system through an autonomous behavior
of the individuals. An independent spinson is immune to the influence of its neighbors and the global field (advertising, mass media). This kind of behavior can also arise if two options (e.g. an old and a new, innovative product) offer both advantages and disadvantages and these advantages and disadvantages are not clearly comparable \cite{bou:bou:03}. In our simulations, at each time step an agent takes a purely random opinion with probability $p$, independently of the neighboring spinsons and product features (i.e. the global field).
\item 
\textit{Conformity} is the effect of a unanimous opinion of a group of neighbors on a spinson that does not behave independently (at a given time step). The nature of these interactions is motivated by the psychological observations of the social impact dating back to Asch \cite{asc:55}: if a group of spinson's neighbors unanimously shares an opinion, the spinson will also accept it. This type of social influence was originally introduced to agent-based models by Sznajd-Weron and Sznajd \cite{s-w:szn:02} and is often referred to as the `Sznajd model' \cite{sta:02}. The original Sznajd model was defined on a 1D lattice and the group of influence was composed of two neighboring spinsons. Stauffer et al. \cite{sta:sou:oli:00} generalized the model to a square lattice and used a $2\times 2$ panel as the group of influence. This became a common approach in 2D models of opinion dynamics and is well motivated by social experiments \cite{s-w:05}. For instance, Asch \cite{asc:55} found that the subjects conformed to a group of four as readily as they did to a larger group, but not so to a smaller group.
\item 
\textit{Advertising} or \emph{mass media} represented by the global external field as in \cite{s-w:wer:03,s-w:wer:wlo:08}, with the strength of the field $h\in [0,1]$ depending on the intensity of advertising and the product features (in a non-specified way). Within our model, with probability $h$ the external field changes the state of a spinson to adopted, if the spinson (i) did not behave independently and (ii) did not yield to group pressure (the group was not unanimous). 
\end{itemize}

Because the focus of this paper is innovation diffusion, at the initial stage of simulations all spinsons are down (i.e., $-1$, $\downarrow$, customers of the old product). If we denote by $c_t$ the ratio (or concentration) at time $t$ of adopted spinsons $N_{\uparrow}(t)$ to the overall number of spinsons in the system $N$, then we can write the initial state as $c_0=0$.

We use a random sequential updating scheme. The time $t$ is measured in Monte Carlo steps (MCS), which consist of $N$ elementary time steps $dt=1/N$. Like in \cite{prz:s-w:wer:13}, each of the elementary time steps is composed of five simulation substeps:
\begin{description}
\item[Substep \#1:]
Randomly choose one spinson $S_i$, which may change its orientation in this time step.
\item[Substep \#2:]
Check if spinson $S_i$ will act independently (with probability $p$) or not (with probability $1-p$). This is implemented by randomly drawing a uniformly distributed number $r\sim U[0,1]$ and checking if $r<p$ (then the spinson acts independently $\rightarrow$ goto \#3) or not ($\rightarrow$ goto \#4).
\item[Substep \#3:]
With probability $f$, after \cite{nyc:s-w:13} called \emph{flexibility}, the chosen spinson changes into the opposite state. To check this, randomly draw a uniformly distributed number $r\sim U[0,1]$. If $r<f$ then $S_i(t+dt)=-S_i(t)$ else $S_i(t+dt)=S_i(t)$. Increase time $t \rightarrow t+dt$ and goto \#1. Note that in this simulation study, without loss of generality, we set $f=0.1$. See \cite{prz:s-w:wer:13} for a discussion of the scaling of model parameters.
\item[Substep \#4:]
Randomly choose a group of $l$ neighboring spinsons that will influence $S_i$. If the group unanimously shares an opinion (i.e. all spinsons have the same value) than $S_i$ will yield to group pressure and take the opinion of the group. Now increase time  $t \rightarrow t+dt$ and goto \#1. If the group is not unanimous than goto \#5.
\item[Substep \#5:]
With probability $h$ spinson $S_i$ is responsive to advertising. This is implemented by randomly drawing a uniformly distributed number $r\sim U[0,1]$ and checking if $r<h$. In such a case (and only in such a case) set $S_i(t+dt)=1$. Finally, increase time $t \rightarrow t+dt$ and goto \#1.  
\end{description} 

Up to this point we have not specified the topology of the spinson network. Real world social networks have usually two properties \cite{mil:67}: (i) links among agents are highly clustered, i.e. on average a spinson's neighbors are likely to be connected to each other, and (ii) the average path length is short, i.e. the average number of intermediaries needed to connect any two spinsons remains relatively low. This unique combination of high clustering and short lengths is an immanent feature of small-world networks (SWN), which appear frequently in diverse types of man-made, biological and ecological systems \cite{new:10}. 
Hence, in this paper we analyze the innovation diffusion process on such a network, generated with the Watts-Strogatz algorithm \cite{wat:str:98}. This procedure was designed as the simplest possible algorithm that accounts for clustering while retaining a short average path length. Recall that to generate a Watts-Strogatz network with $N$ nodes and average degree $K$, we start with a 1D lattice with periodic boundary conditions in which each node is connected to its $K$ neighbors and then with probability $\beta$ replace each edge by a randomly chosen edge. 

\begin{figure}
\centering
\includegraphics[width=7cm]{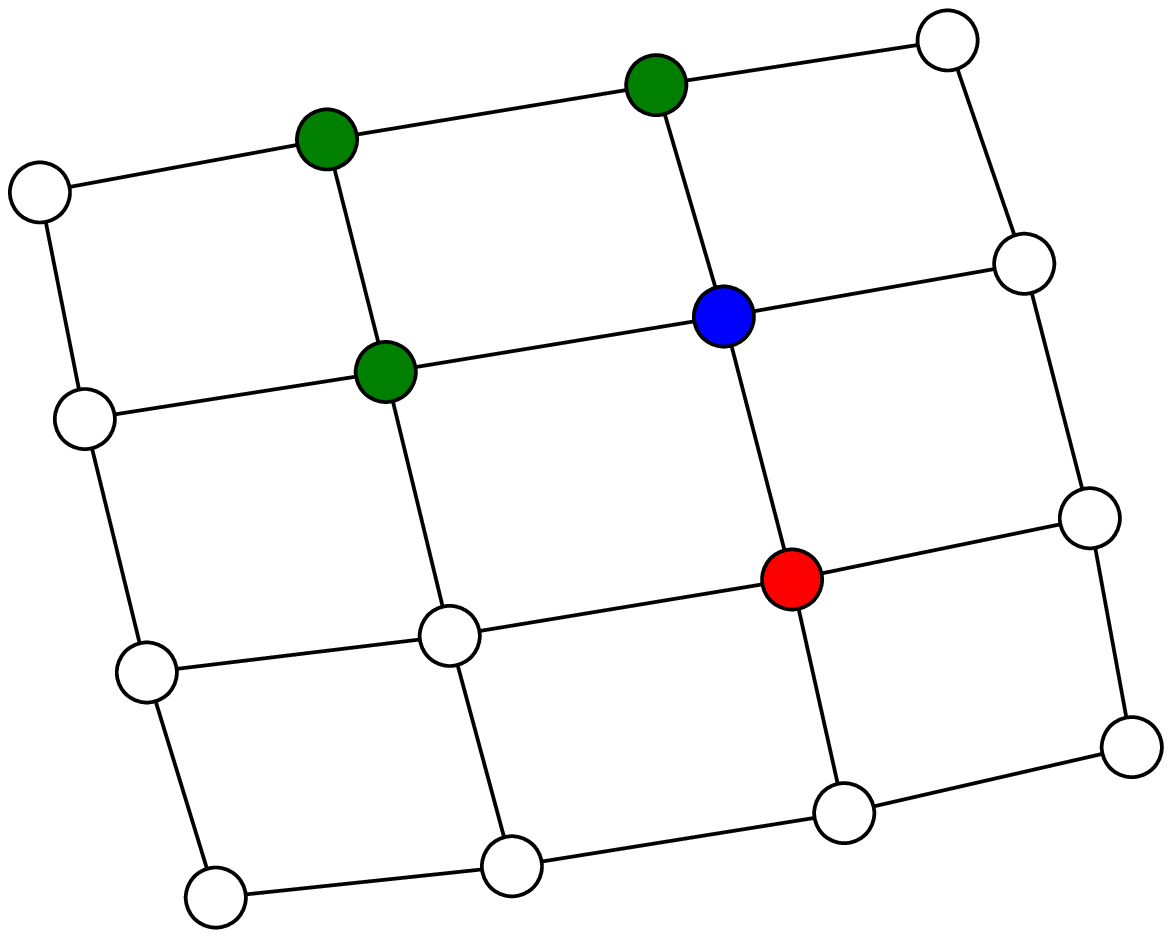}\
\includegraphics[width=7cm]{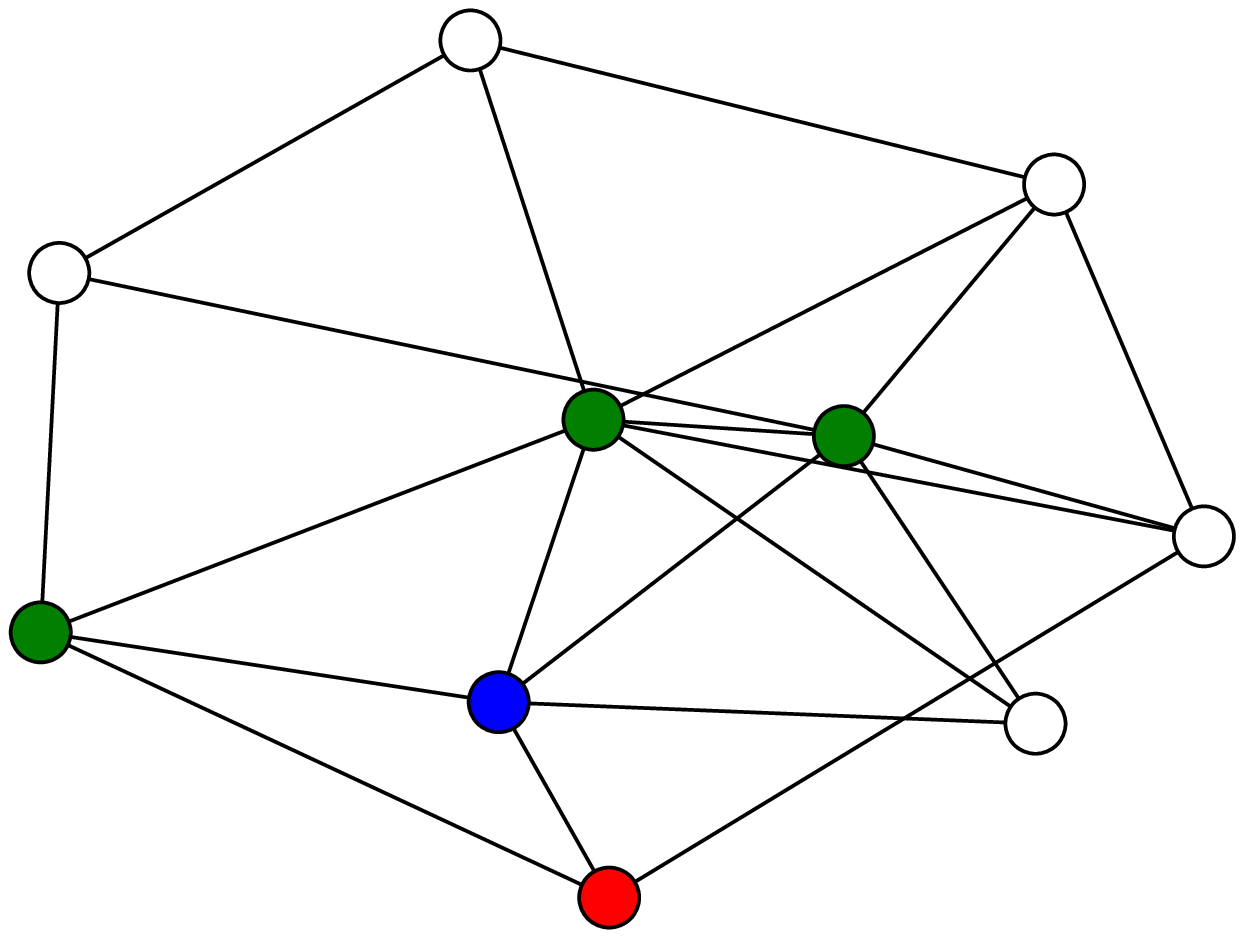}
\caption{An influence group of four spinsons on a complex graph (\emph{right}). For a given spinson ($S_i$, red) we randomly pick one of its neighbors ($S_j$, blue) and then three spinsons from its neighbors (green). As reference, the $2\times 2$ panel typically used on 2D lattices is shown (\emph{left}).}
\label{influence group}
\end{figure}

In Przyby{\l}a et al. \cite{prz:s-w:wer:13} a 2D lattice was used to represent the social system. Conformity was implemented by randomly selecting a spinson,  then one of eight neighboring $2\times 2$ panels and checking if all four spinsons in the panel share the same opinion. Using a small-world network requires redefining the group of influence. Here we use the following scheme (see Fig. \ref{influence group}): 
\begin{enumerate}
\item For a given spinson $S_i$ pick randomly one of its neighbors, i.e. spinsons it is connected to, say $S_j$.
\item Build a list of neighbors of $S_j$ and remove $S_i$ from this list.
\item If the number neighbors of $S_j$ is not less than three, randomly draw three spinsons from the list. They, together with $S_j$, form the influence group. Do nothing if the number neighbors of $S_j$ is less than three, i.e. the influence group is too small to convince $S_i$.  
\end{enumerate}

\section{Results}
\label{sec:results}

\subsection{Class of small-world networks}

As pointed out by Watts and Strogatz \cite{wat:str:98}, the structural properties of a network may be intuitively captured by two quantities -- average path length:
\begin{equation}
L = \frac{1}{N(N-1)}\sum_{i,j}d(v_i,v_j)
\end{equation}
and average cliquishness:
\begin{equation}
C = \frac{1}{N}\sum_i C_i,
\end{equation}
where $N$ is the number of nodes in the network, $d(v_i,v_j)$ denotes the shortest path between nodes $v_i$ and $v_j$, and $C_i$ is the local clustering coefficient of node $v_i$, i.e. the proportion of links between the nodes in its neighborhood and the number of links that could possibly exist between them. 

\begin{figure}
\centering
\includegraphics[width=7cm]{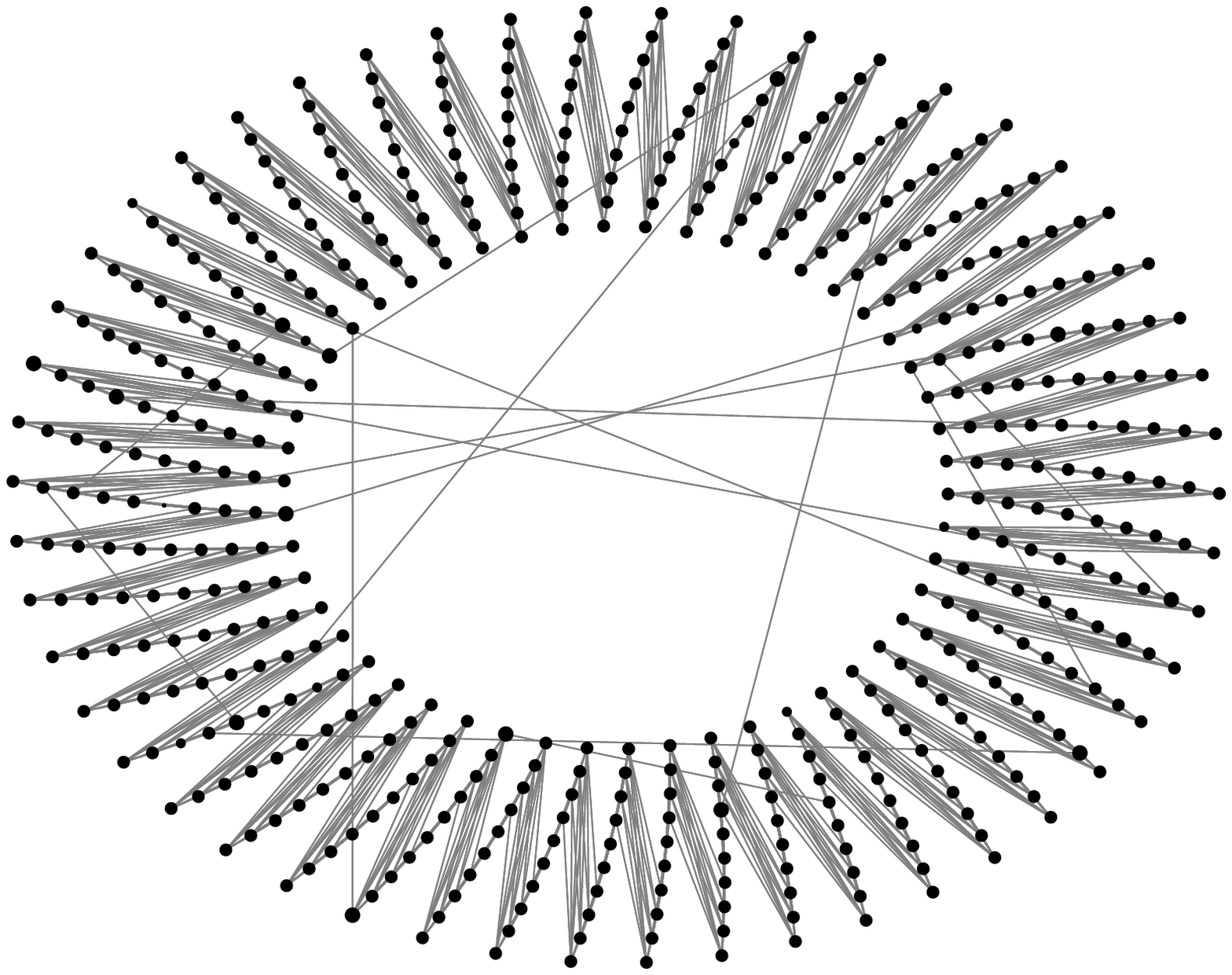}
\includegraphics[width=7cm]{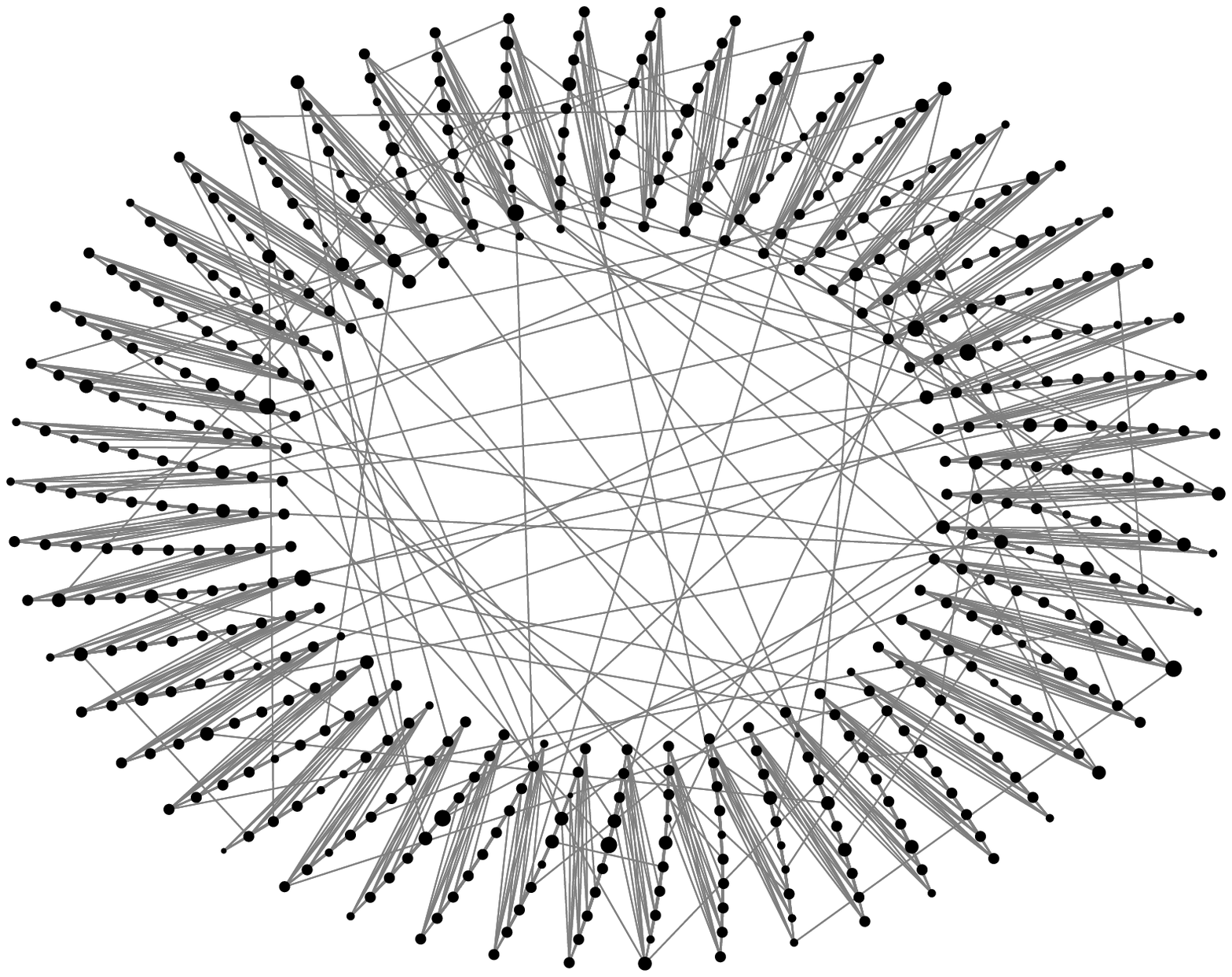}
\caption{Small-world networks of size $N=500$ and mean degree $K=8$ for two values of rewiring: $\beta=0.01$ (\emph{left}) and $\beta=0.05$ (\emph{right}).
\label{small-world network examples}}
\end{figure}

Tuning the rewiring probability $\beta$ in the Watts-Strogatz algorithm allows us to vary the topology of the network from completely regular ($\beta=0$, both $C$ and $L$ high), through intermediate states ($0<\beta<1$), to completely random ($\beta=1$, both $C$ and $L$ low). Among the intermediate states, there is a small interval of rewiring probabilities $0.01<\beta<0.1$ characterized by $L(\beta)\simeq L(1)$ and $C(\beta)\gg C(1)$, i.e. a short path length and high clustering, respectively \cite[Fig. 2]{cow:jon:04}. This interval constitutes the small-world region in the space of networks generated by the Watts-Strogatz method. In Figure \ref{small-world network examples} we plot two sample small-world networks generated by the algorithm for two different values of $\beta$. Note that for $\beta=0.01$ the topology of the network resembles a regular grid. However, already in this case there exist several `shortcuts' in the system that significantly decrease the average path length. 

It should be emphasized that there exists a more quantitative measure of `small-world-ness' introduced by Humphries and Gurney \cite{hum:gur:08}. However, throughout this paper we will use the intuitive distinction between different network topologies based on the values of $C$ and $L$.

\subsection{Time evolution of the system}

\begin{figure}
\centering
\includegraphics[width=14cm]{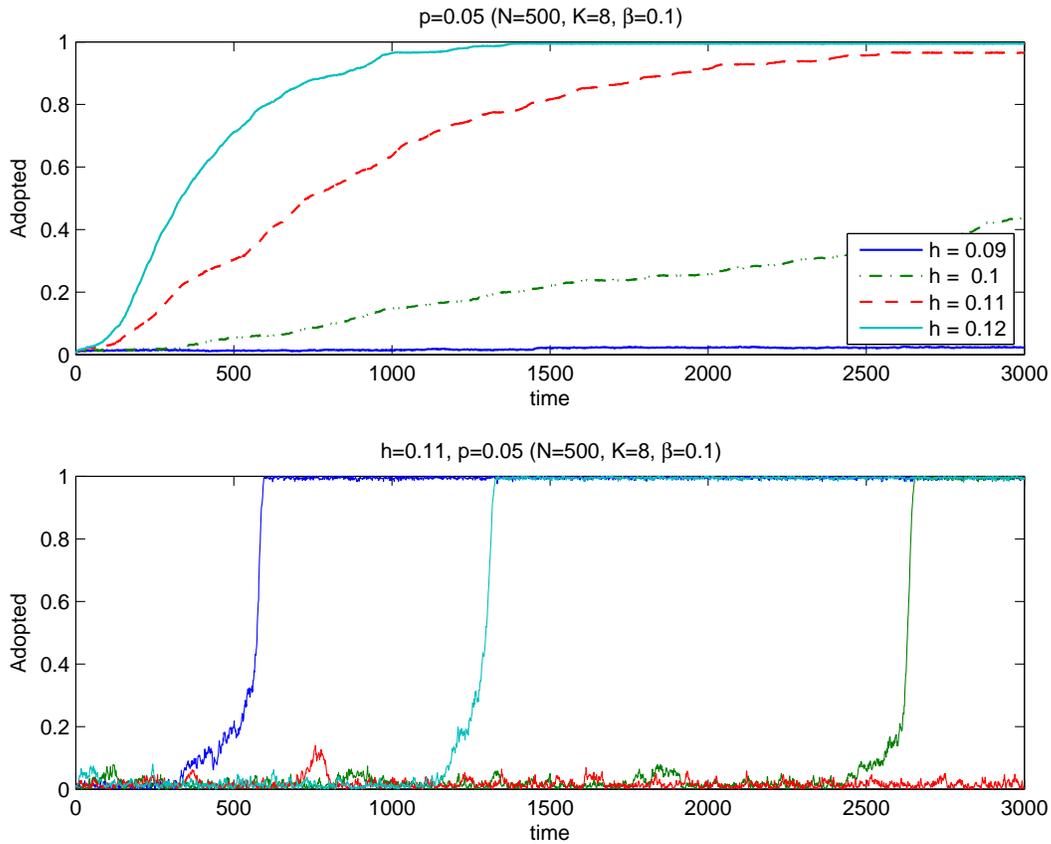}
\caption{\emph{Top panel}: Trajectories representing the time evolution of the average number of adopted spinsons on a small-world network for different external field levels ($h=0.09, 0.10, 0.11, 0.12$) and one independence level ($p=0.05$) for a small-world network with $N=500$ spinsons, mean degree $K=8$ and rewiring probability $\beta =0.1$. Averaging was done over 100 independent runs. 
\emph{Bottom panel}: Four sample trajectories for $h=0.11$. Note the `take-off' phase and the $S$-curve so typical for innovation diffusion. Compare with the average over 100 trajectories in the upper panel (dashed red).
\label{time evolution average}}
\end{figure}

Let us first analyze the time evolution of the average (over 100 independent runs) concentration of adopted spinsons. Trajectories for different external field or advertising levels ($h=0.09, 0.10, 0.11, 0.12$) and one independence level ($p=0.05$) are plotted in the top panel of Fig. \ref{time evolution average} for a small-world network with $N=500$, $K=8$ and $\beta =0.1$.
If we look at individual trajectories for a given set of parameters, see the bottom panel, it turns out that the `take-off' time may vary significantly from run to run. Such a high volatility usually indicates that the system is near the threshold value of $h$, below which the innovation fails and above which it spreads or diffuses in the market \cite{prz:s-w:wer:13}. In what follows we will analyze this issue more deeply. 

In the bottom panel of Fig. \ref{time evolution average}, we see that after a `take-off' phase there is a period of rapid adoptions, followed by a saturation. Thus, our model produces the familiar $S$-curve, in agreement with the empirical studies on the diffusion of innovation \cite{rog:03,kie:gun:stu:wak:12}. Observe in the top panel that the average length of the `take-off' interval depends strongly on the level of advertising. Higher values of $h$ (with other parameters fixed) lead to shorter `take-off' times. In other words, an innovation penetrates the market quicker if it is exposed to more intensive advertising, a conclusion which was to be expected.

\subsection{Phase diagrams}

\begin{figure}
\centering
\includegraphics[scale=0.4]{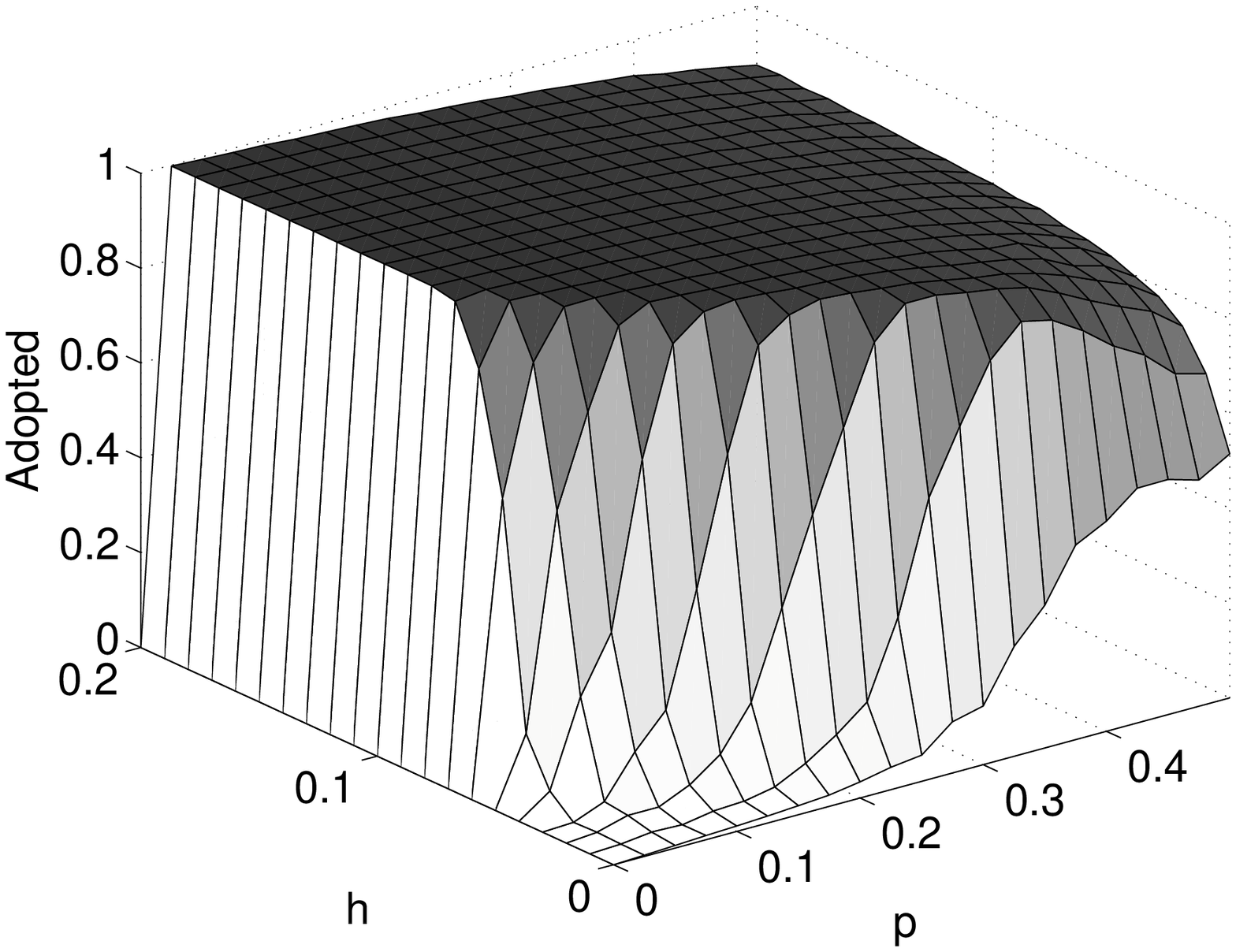}
\includegraphics[scale=0.4]{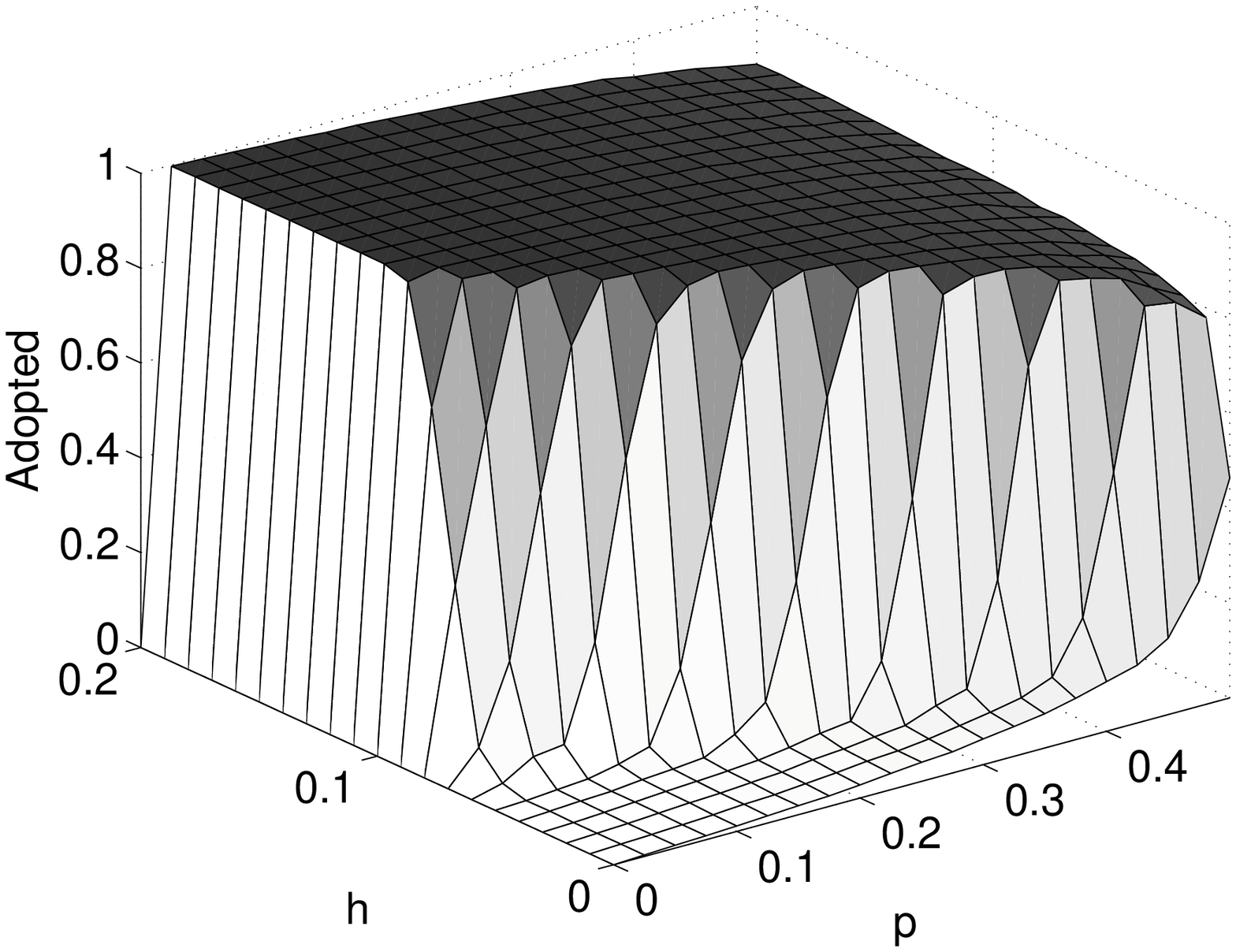}
\includegraphics[scale=0.4]{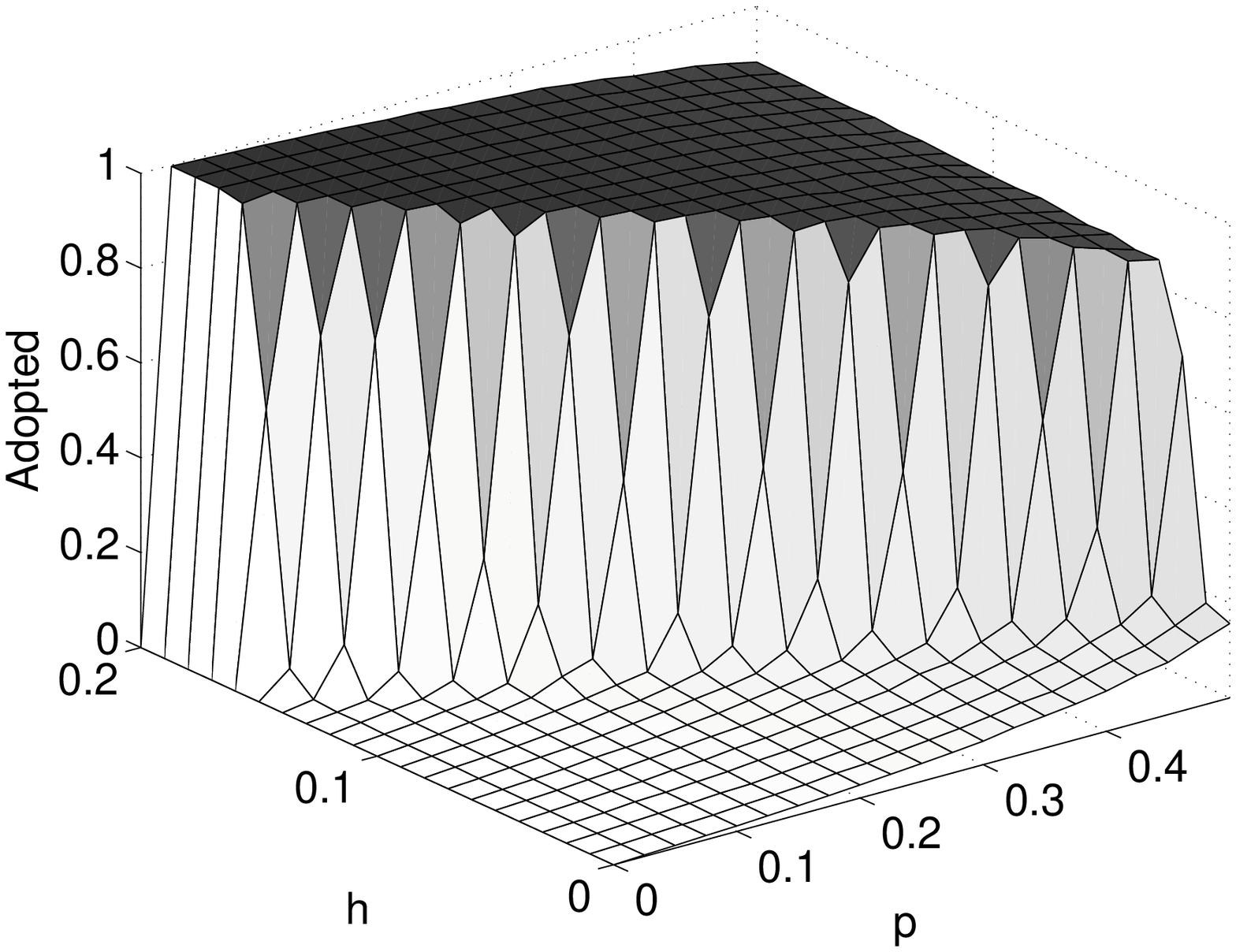}
\includegraphics[scale=0.4]{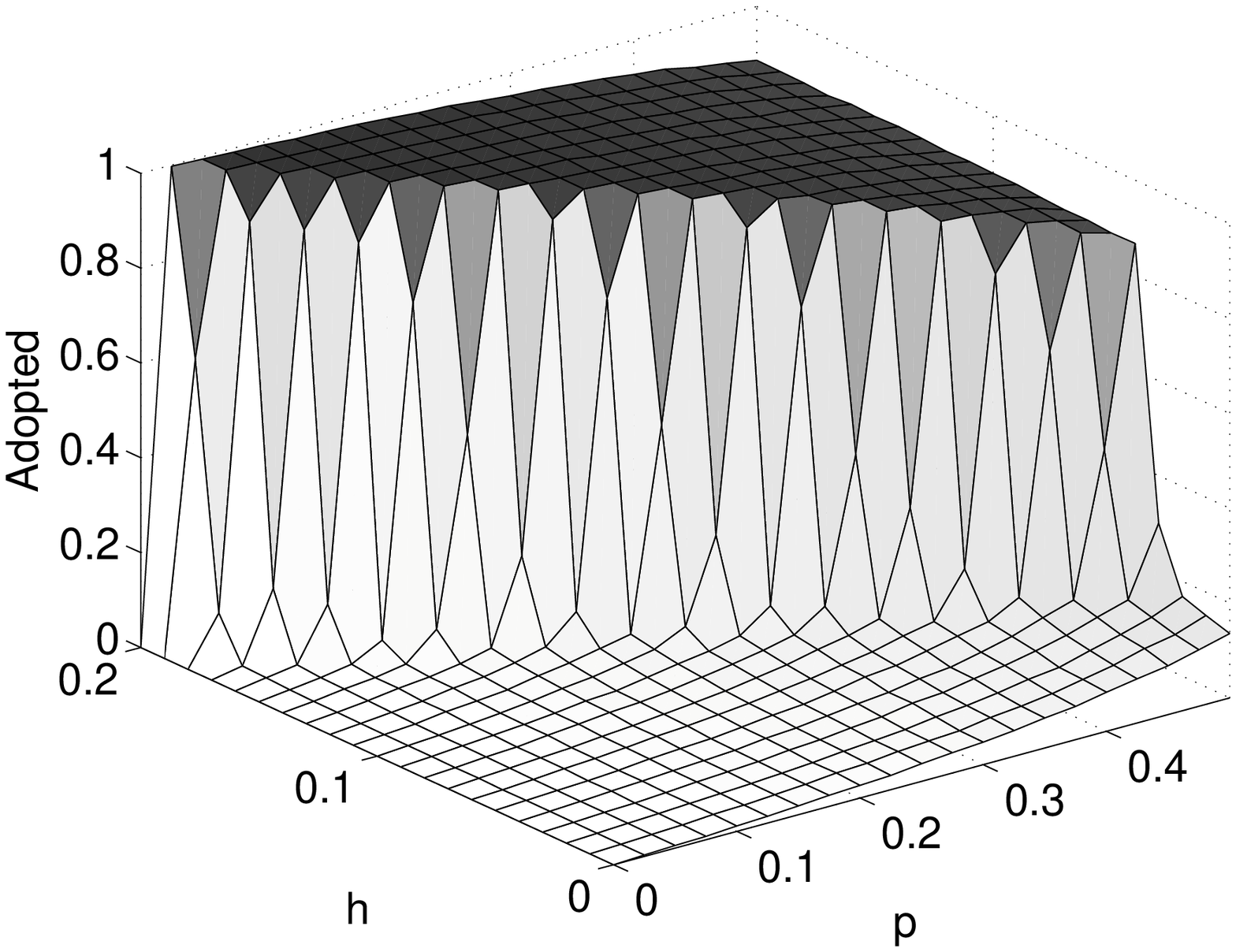}\\
\caption{The ratio of adopted after time $\tau=3000$ MCS as a function of independence $p$ and the external field $h$ for the Watts-Strogatz networks of size $N=500$, mean degree $K=8$ and different rewiring probabilities: $\beta=0$ (regular grid; \emph{top left}), $\beta=0.05$ (small-world; \emph{top right}), $\beta =0.2$ (\emph{bottom left}) and $\beta =1.0$ (random graph; \emph{bottom right}).
\label{phase diagrams}}
\end{figure}

In Figure \ref{phase diagrams}, phase diagrams of the system in the parameter space $(h,p)$ are shown for Watts-Strogatz networks with $N=500$, $K=8$ and four different rewiring probabilities: $\beta=0$, $0.05$, $0.2$ and $1.0$. The first  value of $\beta$ corresponds to a regular grid (no rewiring), the second to a  small-world network with high clustering and short path lengths, the third is an intermediate state between small-world and random networks, and the fourth  corresponds to a random graph with low clustering and a short path length.

In order to generate the diagrams, for each parameter set we measured the average number of adopted spinsons after time $\tau=3000$ MCS. This particular value of $\tau$ turned out to be a reasonable choice, because higher values did not qualitatively impact the results. In other words, if the innovation failed in the first 3000 MCS, it was highly unlikely that it would spread in the market after that time. 

We see that in all cases the innovation fails for small values of $h$ and $p$, since the number of adopted agents is nearly zero at the end of the simulation. However, the system undergoes a phase transition: there exist threshold values of both $h$ and $p$, above which almost the whole population is in the adopted state.
Note that the area of the adopted phase (dark areas in Fig. \ref{phase diagrams}) decreases with higher values of $\beta$. This result might look counter intuitive, because higher $\beta$'s mean a shorter path length, which is typically associated with rapid diffusion. However, the larger the $\beta$ the smaller the cliquishness in the system. It seems that the interplay between the degree of cliquishness and the path length is crucial for a success or failure of the innovation.

\subsection{Does network topology really matter?}

\begin{figure}
\centering
\includegraphics[width=14cm]{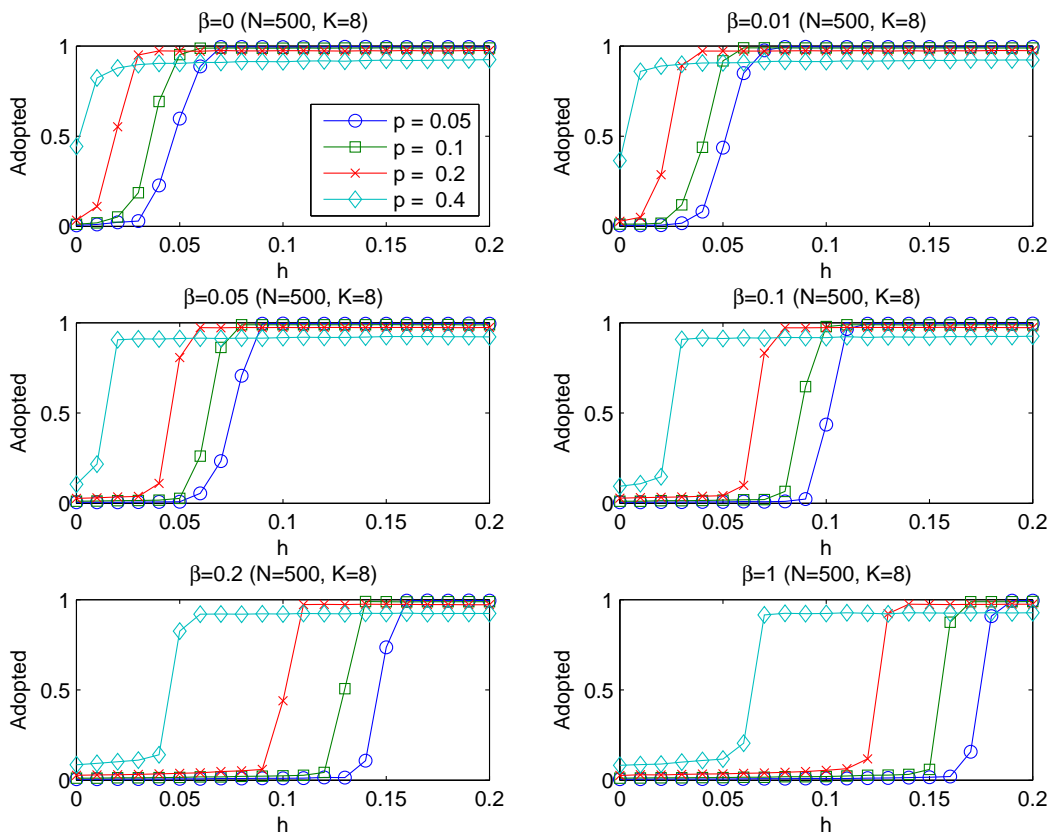}
\caption{Dependence between the ratio of adopted and the external field $h$ on  networks of size $N=500$ and mean degree $K=8$. Each subplot corresponds to a different value of the rewiring probability $\beta$. Note that increasing the randomness of the network (i.e. increasing $\beta$) increases the threshold value of the external field $h^*$, which means that the diffusion of innovation is more difficult in more random societies.
\label{dependence on beta}}
\end{figure}

So far our findings indicate that the actual topology of the Watts-Strogatz network can have a significant influence on the diffusion of innovation. To elaborate on this issue let us consider cross-sections obtained by cutting through the phase diagrams in Fig. \ref{phase diagrams} for fixed values of $p$. The resulting plots are presented in Fig. \ref{dependence on beta}. We see that the threshold values of $h^*$, below which the innovation fails, increase with $\beta$. Thus destroying high clustering by rewiring connections between spinsons is more important for the diffusion of innovation than simultaneous shortening of the path length (both taking place with increasing $\beta$).

\begin{figure}
\centering
\includegraphics[width=14cm]{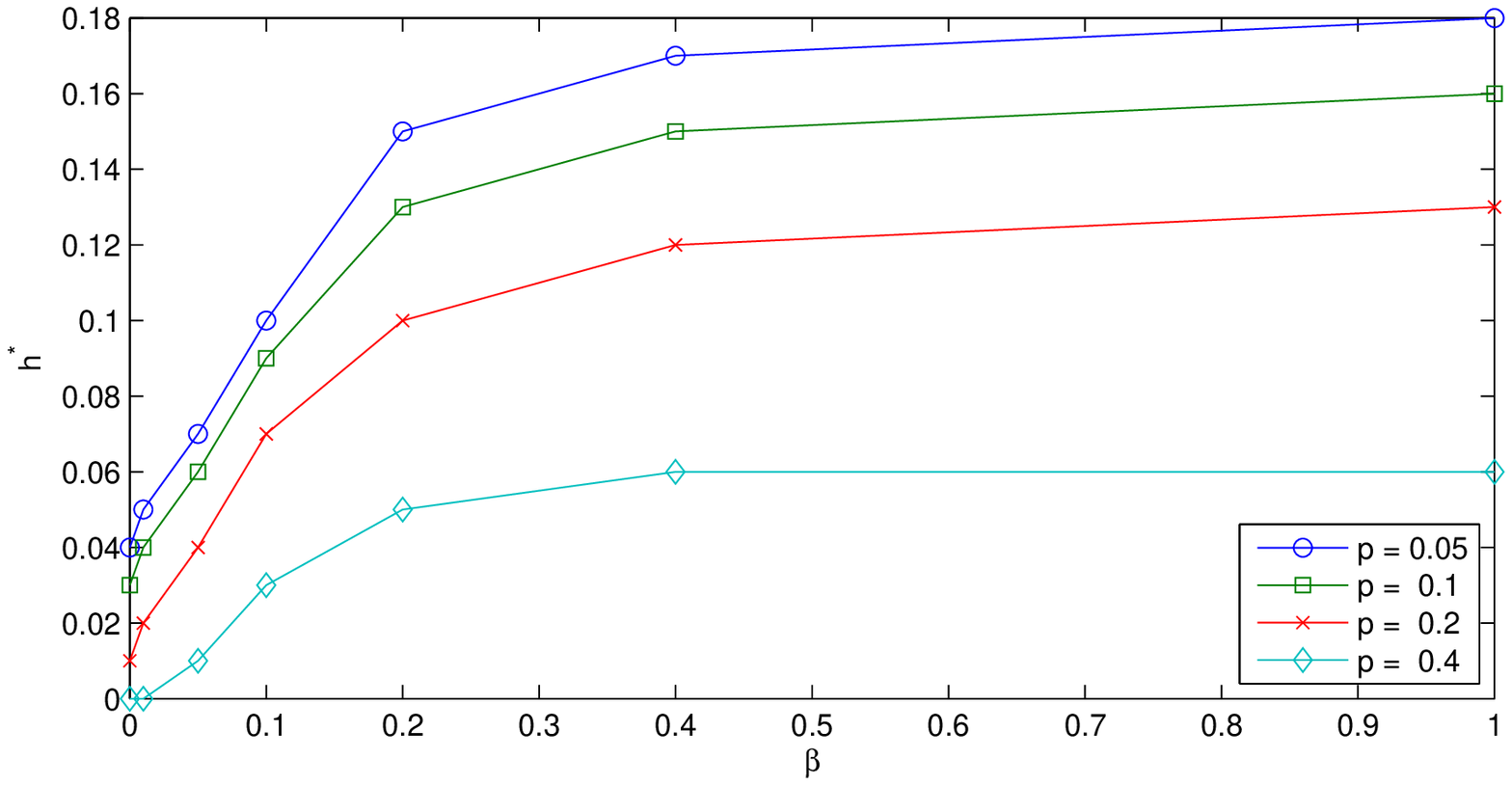}
\caption{Dependence between threshold value $h^*$ of the external field and rewiring probability $\beta$ for the Watts-Strogatz networks of size $N=1000$ and mean degree $K=8$.
\label{h vs beta}}
\end{figure}

It is interesting to note that the dependence of the threshold values $h^*$ on $\beta$ is relatively strong in the small-world class of Watts-Strogatz networks (middle panels in Fig. \ref{dependence on beta}, i.e. for $\beta=0.05$ and $0.1$) and saturates for networks with more randomness (bottom panels in Fig. \ref{dependence on beta}). This result is presented in more detail in Fig. \ref{h vs beta}. Note that in this plot twice larger networks ($N=1000$) are used to obtain smoother curves; for a discussion of finite size effects see Section \ref{ssec:Finite size effects}. 
Aside from the shift in threshold values, the results in Figs.~\ref{dependence on beta} and~\ref{h vs beta}  are qualitatively very similar. In all cases increasing of independence $p$ leads to a shift of the threshold towards smaller values of $h$. This is in agreement with our expectations, because higher independence means more early adopters which are necessary for the innovation to take off. Thus it seems that one may abstract away from the underlying network topology if looking only for qualitative behavior, but the topology matters if one is interested in a quantitative description of the system.

\subsection{Network density}

\begin{figure}
\centering
\includegraphics[width=14cm]{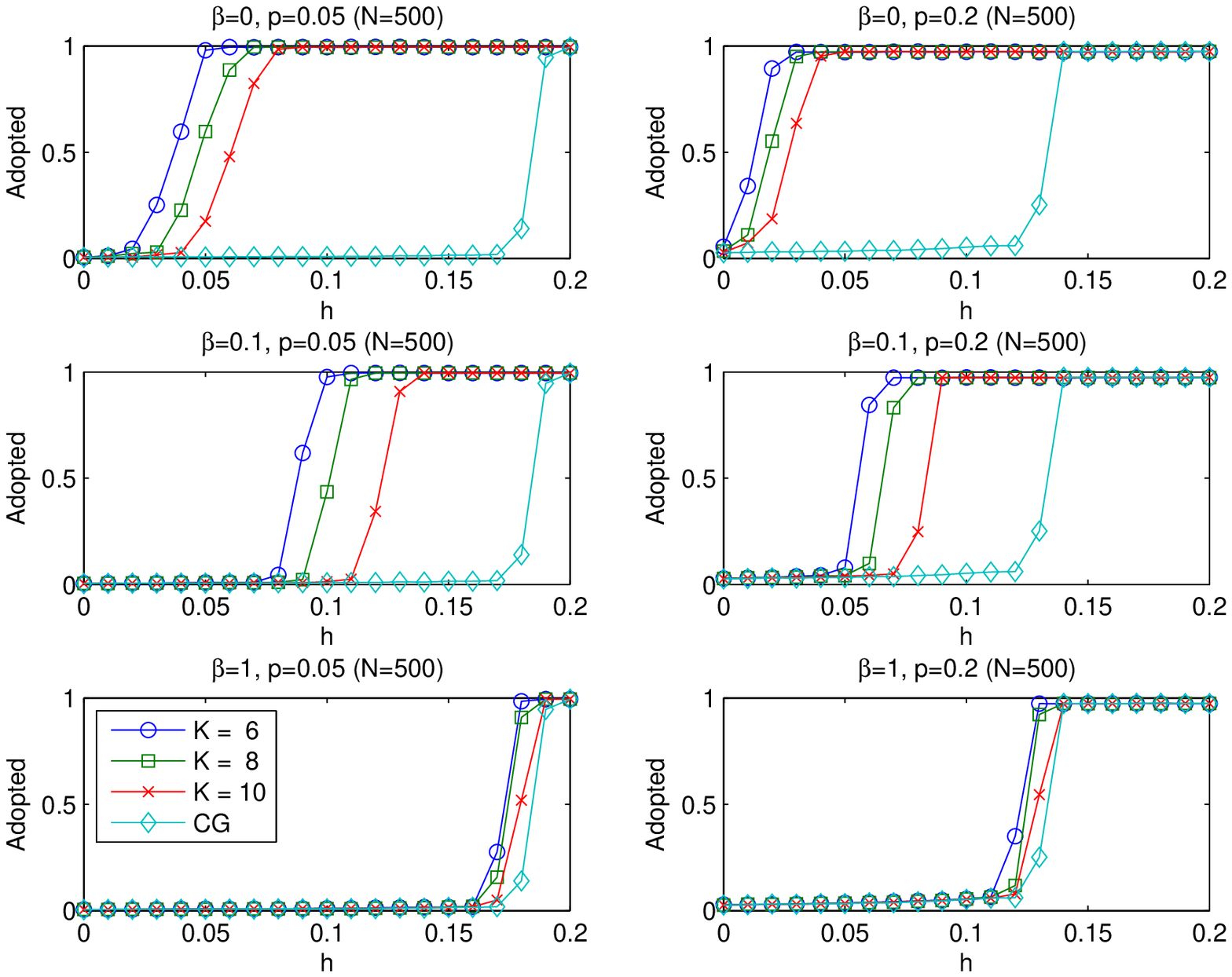}\\
\caption{Dependence between the ratio of adopted and external field $h$ for three Watts-Strogatz networks and a complete graph of size $N=500$. Each subplot corresponds to different values of rewiring probability $\beta$ and independence $p$. In this figure we clearly see how the diffusion of innovation is influenced by the density of the network which corresponds to the mean degree $K$: an innovation is more likely to spread in a sparse network (low $K$) and the influence of $K$ decreases with increasing randomness ($\beta$).
\label{dependence on k}}
\end{figure}

Besides analyzing the impact of $\beta$, it is also interesting to gain insight into the role of the second parameter in the Watts-Strogatz algorithm, the average degree $K$. Since $K$ determines the number of links in the network, it is closely related to the density of the network, i.e. the fraction of links present over all possible links. The final concentration of adopted agents as a function of the external field is shown in Fig. \ref{dependence on k} for different values of $K$. Again, $\beta$ divides the Watts-Strogatz networks into three classes: regular (top panels), small-world (middle panels) and random (bottom panels). For the sake of completeness in Fig. \ref{dependence on k} we also plot the results for a complete graph. Note that the latter corresponds to a small-world network with $K=N-1$ and rewiring probability $\beta=0$.

Surprisingly, at least at first sight, the threshold values for external field $h$ increase with $K$ for small values of $\beta$. Moreover, the influence of $K$ decreases with increasing randomness of the network and for a completely random graph ($\beta=1$) the results are essentially independent of $K$. This result implies that the diffusion of innovation is more difficult in dense societies with a high degree of randomness. The explanation may go along the following line: in a sparse network (low $K$) a relatively small number of adopters may destroy the unanimity of most possible influence groups and the system is more  susceptible to the external field.

\begin{figure}
\centering
\includegraphics[width=14cm]{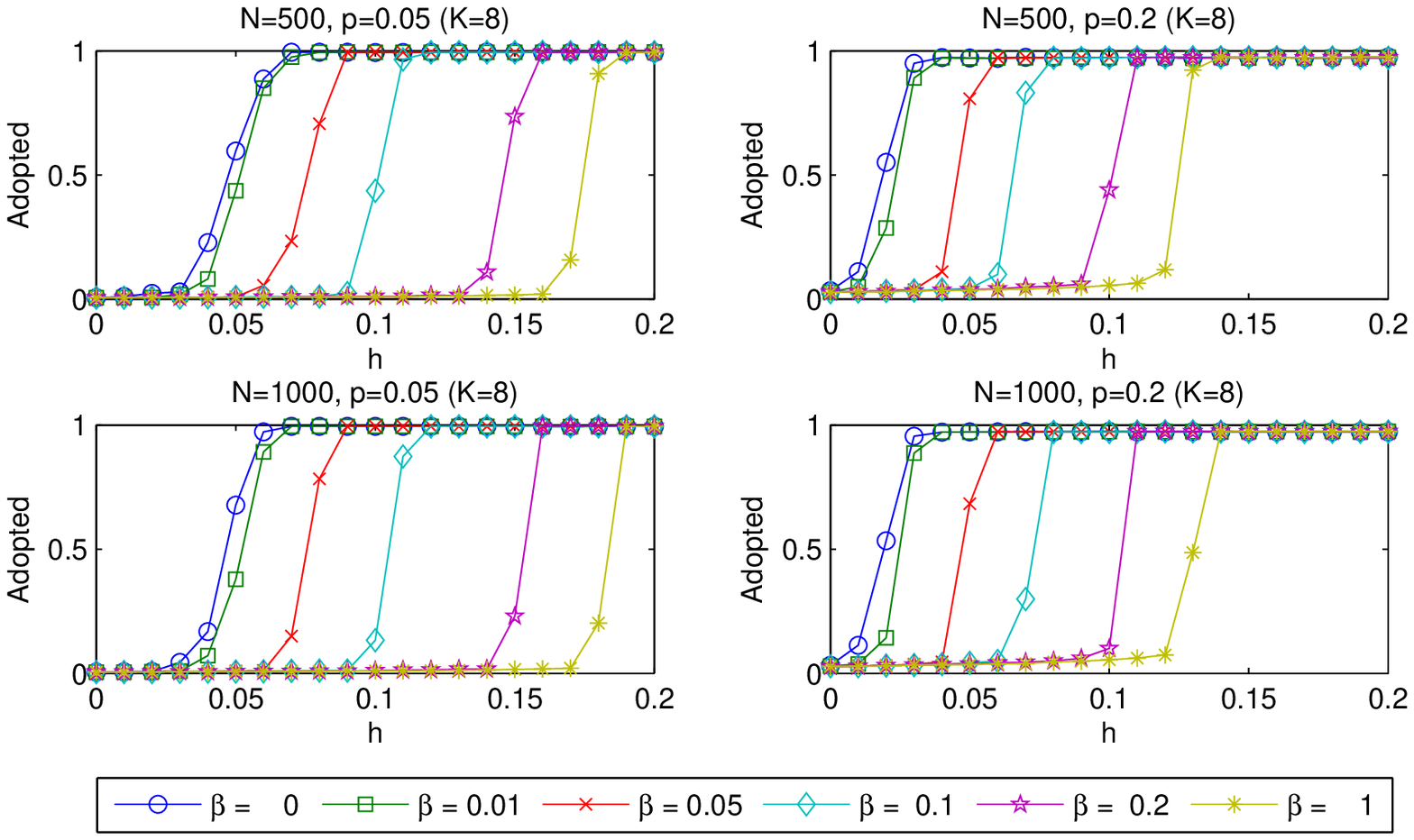}
\caption{Dependence between the ratio of adopted and external field $h$ on Watts-Strogatz networks for two values of independence $p$ and two system sizes $N$. Note that increasing network randomness (i.e. $\beta$) increases the threshold value of the external field. Note also that threshold value $h^*$ increases with the size of the system (compare the upper with the lower panels).
\label{dependence on N} }
\end{figure}

\subsection{Finite size effects}
\label{ssec:Finite size effects}

In Figure \ref{dependence on N} we illustrate the dependence of threshold $h^*$ on the size of the system. For all values of $\beta$ the threshold increases with $N$, i.e. the diffusion of innovation is hindered by the number of agents.
To explain this result, at least qualitatively, let us consider a simplified version of our model with the size of the influence group equal to 1 (see Section \ref{sec:model} for model details) and the complete graph as the underlying network topology. Let us assume that at time $\tau$, the first adopted spinson appeared in the system (due to independence) and that in the following Monte Carlo event we have chosen a not-adopted spinson. In order to find its influence group we just have to pick one of the other spinsons at random (all spinsons are connected). Thus, the probabilities of finding an adopted influence group and a not-adopted one are $1/(N-1)$ and $(N-2)/(N-1)$, respectively. We see that the larger the system size the smaller the probability of finding an adopted group and changing the opinion of the spinson under consideration. This is why the diffusion of innovation is more difficult in larger systems.

\section{Conclusions}
\label{sec:conclusions}

In this paper we have studied the role of network topology within the innovation diffusion model introduced recently in \cite{prz:s-w:wer:13} and applied to a real-life problem of the adoption of dynamic electricity tariffs in \cite{kow:mac:s-w:wer:13}. In both cited papers the model was studied on a square lattice, despite empirical evidence showing that the topology of a real-world social network cannot be properly described by a regular graph. However, the role of network topology in innovation diffusion models is still unclear, most likely due to the crucial role of social influence in such problems \cite{rog:03,nol:sch:cia:gol:gri:08,alc:11,ayr:ras:shi:13}. This may sound surprising given the fact that the influence of network topology on the diffusion of knowledge is a well known phenomenon \cite{lin:li:10}. Indeed papers that study diffusion of knowledge show that a short path increases the diffusion power of the system in a sense that the knowledge will move quicker to different parts of the graph \cite{cow:jon:04}. Therefore knowledge spreading is much harder in a regular graph than in a small-world network or a random graph. On the other hand, the role of network topology in the  diffusion of innovation is highly unclear and within different models completely different results have been obtained, as discussed in the Introduction and in \cite{wat:dod:07,koc:kun:08,boh:cal:zha:10,peg:ren:mar:12}. Therefore the focal point of this study was the role of network topology on innovation diffusion. A natural second step was seeing how our results compare to the results obtained by other authors and whether our analysis can shed more light on the problem.

To investigate the role of network topology we have decided to use the Watts-Strogatz model \cite{wat:str:98}. For several reasons. Firstly, because this model -- depending on the selected parameters -- allows to study various topologies: 
\begin{itemize}
\item a regular graph (for $\beta=0$ and any value of $K$), 
\item a complete graph (for $\beta=0$ and $K=N-1$),
\item a random graph (for $\beta=1$ and any value of $K$), 
\item a small-world network (for $0.01 <\beta <0.1$ and any value of $K$), 
\item and other topologies that are more random than small-world but not completely random.
\end{itemize}
Secondly, because by changing $K$ we were able to investigate the role of the network density, which is an important characteristic from a social point of view. Thirdly, the small-world topology -- which can be obtained within Watts-Strogatz model -- adequately describes most features of real-world societies \cite{wat:str:98}. 

As we have discussed in the Introduction, the first step in the diffusion of innovation is gaining knowledge about the innovation. From this point of view rewiring a regular network should help the innovation to diffuse. However, in our model we assume that such information is widely available through the actions of mass media, advertising, etc. Therefore we focus on the second stage, the so-called persuasion stage \cite{rog:03}, for which social influence is the most important. Because social influence is most powerful in the presence of unanimity and decays dramatically with physical distance \cite{lat:etal:95,bat:mar:now:06,bru:rei:ryf:07} it is not obvious if rewiring the network is desired at this stage of real-life innovation diffusion processes. 

Within our model we have found that there is a critical value of the external field, $h^*$, above which the innovation diffuses and below which it fails. This critical value depends on the level of independence $p$ -- the higher the independence the smaller the critical field needed for the innovation to diffuse. An analogous result has been obtained for a square lattice \cite{prz:s-w:wer:13}. The really new results that have been obtained here are:
\begin{itemize}
\item The critical value of the external field, $h^*(p)$, increases with $\beta$, which means that the innovation diffuses more likely in more regular graphs.
\item Generally, the critical value of the external field increases with $K$, which means that it is harder for the innovation the spread in more dense networks. However, the influence of $K$ decreases with $\beta$ and for $\beta=1$ there is practically no dependence on $K$.
\item Results obtained for the complete graph are identical with those obtained for the random graph (for any value of $K$), which can be understood using the mean field approach \cite{prz:s-w:wer:13}.
\end{itemize}

We have mentioned in the Introduction that one can find different and at times seemingly contradictory results comparing agent-based studies of the diffusion of innovation. However, looking at these results again, and having conducted our study, we may conclude that they are in fact consistent and allow to predict the role of network topology in various situations. For example, Bohlmann et. al. \cite{boh:cal:zha:10} have investigated the role of topology in the threshold model, which postulates that, for an agent to adopt, the proportion of its local network that has already adopted must reach a minimum level (threshold) and have found that more clustered networks appear more likely to diffuse under high adoption thresholds. On random graphs the diffusion failed completely for higher thresholds. This agrees with our results that the diffusion of innovation is easier on more regular graphs (i.e. with a higher clustering coefficient). Of course, we have not been investigating the threshold model but our model is somehow similar. We need an unanimous group to influence a neighboring spinson, which corresponds to the maximum value of the threshold (i.e. equal to 1). On the other hand, in our case we do not consider all neighbors of a given spinson but always a panel of four and this makes the models different. 

Another consistent result has been obtained by Pegoretti et al. \cite{peg:ren:mar:12}. They have shown that with perfect information, the innovation diffuses faster when the network is completely random, but with imperfect information the fastest pace is reached in small-world networks. We should realize that the situation when an innovation is entering the market is one of high uncertainty, which is related to imperfect information. In such situations people do not behave rationally and social influence is extremely important \cite{smi:hog:mar:ter:07}. Therefore, we could conclude that in the case of uncertainty, which is particularly high for innovations connected to public health programs or ecological campaigns \cite{rog:03}, a more clustered network will help the diffusion. On a random graph the diffusion will be more difficult than in a small-world network or a regular lattice. On the other hand, when social influence is less important (i.e. in the case of perfect information), 
a shorter path will help the innovation to spread in the society and -- as a result -- the diffusion will be easiest on a random graph.

Finally, concerning the role of network density, we have found that it is very hard to adopt an innovation in a dense society. This, however, can be easily understood because in dense societies social pressure is very high and it is very difficult to adopt a new idea. This result could explain many puzzles in the field of innovation diffusion, like the failure of the water-boiling campaign conducted in Los Molinas and many other health campaigns in small villages, where almost everyone knows everyone \cite{rog:03}.

\section*{Acknowledgments}

This work was partially supported by funds from the National Science Centre (NCN) through grant no. 2011/01/B/ST3/00727.

\end{document}